\documentclass{article}

    \PassOptionsToPackage{numbers, compress}{natbib}

    \usepackage[preprint]{neurips_2024}

\usepackage[utf8]{inputenc} 
\usepackage[english]{babel}

\usepackage[T1]{fontenc}    
\usepackage{hyperref}       
\hypersetup{
    colorlinks=true, 
    linkcolor=blue, 
    urlcolor=blue,
    citecolor=blue 
}
\usepackage{url}            
\usepackage{booktabs}       
\usepackage{amsfonts}       
\usepackage{nicefrac}       
\usepackage{microtype}      
\usepackage{xcolor}         

\usepackage{amsmath}    

\usepackage{graphicx}
\usepackage{booktabs}   
\usepackage{siunitx}    
\title{CoGenAV: Versatile Audio-Visual Representation Learning via Contrastive-Generative Synchronization}

\author{Detao Bai$^{1}$,Xihan Wei$^{1}$,Liefeng Bo$^{1}$,Zhiheng Ma$^{2}$\\
$^{1}$Tongyi Lab, Alibaba Group\\
$^{2}$ Shenzhen University of Advanced Technology\\
\url{https://github.com/HumanMLLM/CoGenAV}
}

\begin{document}

\maketitle

\begin{abstract}
   The inherent synchronization between a speaker's lip movements, voice, and the underlying linguistic content offers a rich source of information for improving speech processing tasks, especially in challenging conditions where traditional audio-only systems falter. We introduce CoGenAV, a powerful and data-efficient model designed to learn versatile audio-visual representations applicable across a wide range of speech and audio-visual tasks. CoGenAV is trained by optimizing a dual objective derived from natural audio-visual synchrony—contrastive feature alignment and generative text prediction—using only 223 hours of labeled data from the LRS2 dataset. This contrastive-generative synchronization strategy effectively captures fundamental cross-modal correlations. We showcase the effectiveness and versatility of the learned CoGenAV representations on multiple benchmarks. When utilized for Audio-Visual Speech Recognition (AVSR) on LRS2, these representations contribute to achieving a state-of-the-art Word Error Rate (WER) of 1.27. They also enable strong performance in Visual Speech Recognition (VSR) with a WER of 20.5 on LRS2, and significantly improve performance in noisy environments by over 70\%. Furthermore, CoGenAV representations benefit speech reconstruction tasks, boosting performance in Speech Enhancement and Separation, and achieve competitive results in audio-visual synchronization tasks like Active Speaker Detection (ASD).Our code: \url{https://github.com/HumanMLLM/CoGenAV} 
\end{abstract}

\section{Introduction}

Human communication inherently leverages multiple modalities, with speech perception naturally integrating auditory signals and visual cues like lip movements. This audio-visual synchrony is fundamental; the visual stream (lip articulation) offers information complementary to the audio, particularly valuable when the acoustic signal is corrupted by noise, distorted, or originates from overlapping speakers. While recent Automatic Speech Recognition (ASR) systems, including large models like Whisper \cite{whisper}, SenseVoice \cite{SenseVoice}, and Qwen-Audio \cite{qwen-aduio,qwen2-audio}, have achieved impressive accuracy on clean, well-resourced benchmarks, their performance often degrades significantly in more challenging real-world scenarios. Persistent difficulties remain in handling high ambient noise, and disentangling multi-speaker conversations \cite{varga1993assessment,Auto-avsr,mWhisper-flamingo,XLAVS-R}. 
The robustness of the visual speech signal to acoustic interference suggests that leveraging audio-visual synchrony is a key, yet underexplored, avenue to overcome these fundamental limitations of audio-only ASR.

Motivated by this potential, enhancing speech processing with audio-visual information has become an active research area \cite{av-hubert, Auto-avsr, USR}. While early works demonstrated performance gains by integrating visual cues, they often resulted in task-specific models reliant on large datasets or complex training paradigms \cite{Auto-avsr}. Learning versatile and data-efficient audio-visual representations thus remains a core challenge. Recent attempts adapt large pre-trained ASR models by incorporating visual features \cite{whisperer, Whisper-flamingo}, but typically require fine-tuning the entire large model. This strategy faces significant hurdles, including high computational costs and the risk of catastrophic forgetting \cite{cot_cat}, limiting its practicality and scalability. These limitations underscore the need for alternative approaches to effectively leverage audio-visual synchrony.

Addressing this need, we propose CoGenAV (Contrastive-Generative Audio-Visual model), a novel approach for learning powerful and versatile audio-visual representations directly from the inherent synchrony of speech. CoGenAV is trained by optimizing a dual objective function grounded in this natural consistency. Specifically, using datasets containing synchronized audio, video, and text (such as LRS2 \cite{LRS2}), the model simultaneously learns to: 1) maximize the alignment between audio and visual features via a sequence-to-sequence contrastive loss, capturing fine-grained temporal correspondence, and 2) predict the associated text transcription using a generative log-likelihood loss, thus embedding semantic information. Notably, this combined contrastive-generative synchronization strategy proves highly effective even with moderate amounts of labeled data; our core training utilizes only the 223 hours available in LRS2 \cite{LRS2}, demonstrating significant data efficiency.

The representations learned by CoGenAV, owing to the contrastive-generative synchronization objective and data-efficient training (on 223 hours of labeled data), encapsulate fundamental properties of audio-visual speech. By jointly modeling cross-modal alignment and associated linguistic content, they inherently capture rich temporal and semantic correlations. Furthermore, the intrinsic integration of visual information provides a representation stream naturally resilient to acoustic noise. These core properties suggest their potential for broad applicability across various audio-visual processing challenges where synchrony or acoustic robustness is crucial.

We conducted extensive experiments to validate the effectiveness and versatility of the CoGenAV representations. Demonstrating their strength in transcription tasks, utilizing CoGenAV features enables achieving a WER of 20.5 in VSR on LRS2 \cite{LRS2} (outperforming prior methods \cite{Auto-avsr}) and contributes to a state-of-the-art AVSR WER of 1.27 \cite{Auto-avsr, USR}. Highlighting their robustness, performance in noisy environments (0dB SNR) improves by over 70\% compared to strong audio-only baselines. Underscoring their versatility, these representations also yield substantial gains when used as visual features for speech reconstruction (Enhancement and Separation) and achieve competitive performance in synchronization tasks like Active Speaker Detection on the Talkies dataset \cite{Talkiez}. 

This work's primary contributions are threefold: 1) The proposal of CoGenAV, a model learning powerful audio-visual representations via a dual contrastive-generative synchronization objective. 2) A novel audio-visual sequence-to-sequence contrastive learning approach integrated within CoGenAV to effectively capture fine-grained temporal alignment between modalities. 3) Extensive experimental validation demonstrating the versatility, data efficiency, and state-of-the-art or competitive performance of CoGenAV representations across diverse audio-visual tasks using only moderate training data.

\section{Method}
\begin{figure}
\centering
\includegraphics[width=0.9 \linewidth]{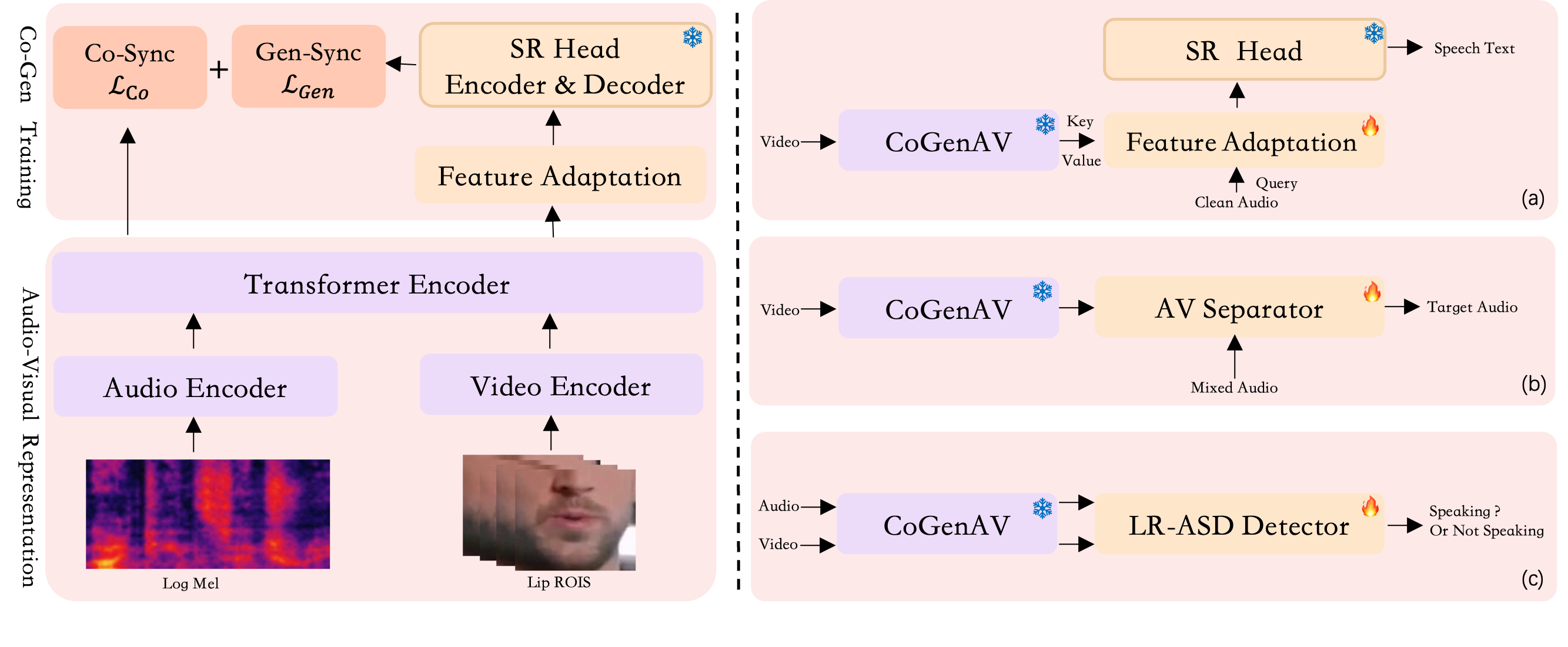}
\caption{\label{fig:archmodel}Our method. Left: Audio-Visual  Representation framework and the Contrastive-Generative Synchronization Training methodology. The SR Head represents the encoder and decoder components of a frozen pre-trained ASR model for Speech Recognition. Right: CoGenAV applied to diverse downstream tasks, (a) CoGenAV for  AVSR; (b) CoGenAV for AVSS and AVSE; (c) CoGenAV for ASD. The blue snowflake represents weights that are frozen and non-trainable.}
\end{figure}

Our architecture is illustrated in Figure~\ref{fig:archmodel}. The left panel depicts the Audio-Visual Feature Representation framework and the Contrastive-Generative Synchronization Training methodology. For generative synchronization, we design a Feature Adaptation Module and employ a frozen pre-trained ASR model as the Speech Recognition (SR) head. The right panel demonstrates the application of CoGenAV to diverse downstream tasks, including Visual Speech Recognition (VSR), Audio-Visual Speech Recognition (AVSR), Audio-Visual Speech Separation (AVSS), Audio-Visual Speech Enhancement (AVSE), and Active Speaker Detection (ASD).

\subsection{Audio-Visual feature  representation}

The audio-visual feature representation shares the same ResNet-3D CNN visual encoder and Transformer encoder as AV-HuBERT \cite{av-hubert}, along with a new audio encoder that ensures frame-synchronous alignment of audio features with visual features while maintaining compatibility with the audio preprocessing methods of pre-trained ASR models.

The Audio Encoder consists of two 3x3 convolutional layers with stride 2 and  GELU activation functions, progressively reducing the temporal dimension by a factor of 4,encoding raw audio Mel-spectrogram features \(a_t \in \mathbb{R}^{4T \times 80}\) into high-dimensional embeddings \(f_t^a \in \mathbb{R}^{T \times D}\).  The visual input comprises \(T\) frames of lip region images \(v_t \in \mathbb{R}^{T \times H \times W \times C}\), processed through a ResNet18-like 3D CNN encoder, resulting in output \(f_t^v \in \mathbb{R}^{T \times D}\). Consequently, audio and visual features \(f_t^a\) and \(f_t^v\) are temporally aligned with the same feature dimensionality.

The CNN-extracted features are then fed into the Transformer Encoder to capture temporal and global relationships. For audio-only input, we obtain \(F_t^a \in \mathbb{R}^{T \times D}\) from \(f_t^a\); for video-only input, we derive \(F_t^v \in \mathbb{R}^{T \times D}\) from \(f_t^v\). For simultaneous audio and video input, we concatenate \(f_t^a\) and \(f_t^v\) before passing them through the Transformer Encoder to obtain audio-visual features \(F_t^{av} \in \mathbb{R}^{T \times D}\). Unlike time-aggregated representations from audio-visual contrastive learning methods such as SyncNet\cite{SyncNet} and LatentSync\cite{LatentSync}, our features retain the temporal dimension \(T\) in the two-dimensional representations \(F_t^a\) and \(F_t^v\). Based on this, we employ a Seq2Seq contrastive learning approach in this paper.

\subsection{Contrastive-Generative synchronization training}

The speech signal and the speaker's lip movements are inherently synchronized and correspond to the same textual expression.\textbf{CoGenAV} establishes tri-modal alignment representations across audio, visual, and text streams through joint training of contrastive synchronization and generative synchronization.Contrastive alignment enforces frame-level audio-visual consistency through cosine similarity maximization, while generative prediction aligns latent features with a frozen ASR model's acoustic-textual.

\subsubsection{\textbf{Contrastive synchronization }}

In contrastive synchronization, employ the CoGenAV Representation to extract audio-only features \(F_t^a \in \mathbb{R}^{T \times D}\) and video-only features \(F_t^v \in \mathbb{R}^{T \times D}\), and align them via a Seq2Seq Contrastive self-supervised approach. The core objective is to enforce the CoGenAV model to output \textbf{cross-modally consistent }visual and audio features while ensuring their \textbf{temporal synchronization}.  

To align audio features \( F^a \in \mathbb{R}^{T \times D} \) and visual features \( F^v \in \mathbb{R}^{T \times D} \), we design a sequence-to-sequence contrastive loss  \( \mathcal{L}_{\text{Co}} \).Instead of computing a single embedding for the entire sequence, we first calculate frame-wise cosine similarities \( \frac{\langle F_{i,t}^a, F_{i,t}^v \rangle}{\|F_{i,t}^a\| \|F_{i,t}^v\|} \) for each time step \(t\).We apply the ReLU activation function to these frame-wise similarities, ensuring non-negativity (\( \in [0, 1] \))and potentially increasing robustness by down-weighting frames with strong negative correlation, which might arise from noise or transient asynchronies. The sequence-level similarity \( \bar{d}_i \) is then computed as the temporal mean of these non-negative frame similarities(Eq.\ref{eq:bar_d}). This approach aims to preserve local temporal information more effectively than methods relying solely on global sequence representations. Finally, we formulate the objective as a binary cross-entropy (BCE) loss  (Eq.\ref{eq:L_Co})between \( \bar{d}_i \in [0, 1] \) and the ground truth  \( y_i \in \{0,1\} \), facilitating direct optimization for synchrony detection. Negative pairs ( \( y_i = 0 \)) are constructed using both cross-speaker samples and same-speaker samples with deliberate temporal misalignment,positive pairs ( \( y_i = 1 \))  are obtained from the same speaker's raw recordings that are strictly time-aligned.The loss \( \mathcal{L}_{\text{Co}} \) is formulated as: 
\begin{gather}
    \bar{d}_i = \frac{1}{T} \sum_{t=1}^T \mathrm{ReLU}\left( \frac{\langle F_{i,t}^a, F_{i,t}^v \rangle}{\|F_{i,t}^a\| \|F_{i,t}^v\|} \right) 
    \label{eq:bar_d}  \\ 
    \mathcal{L}_{\mathrm{Co}} = -\frac{1}{N} \sum_{i=1}^N \left[ y_i \log \bar{d}_i + (1 - y_i) \log(1 - \bar{d}_i) \right]
    \label{eq:L_Co}
\end{gather}
where \( \bar{d}_i \in [0, 1] \) computes the temporal mean of \textbf{frame-wise non-negative cosine similarities}, \( T \) denotes the sequence length about time steps, \( N \) is the batch size, and \( y_i \in \{0,1\} \) indicates alignment labels (1 for synchronized pairs, 0 otherwise).  

The Seq2Seq contrastive learning method preserves temporal locality information, capturing fine-grained alignment and enabling the detection of local alignment relationships between audio and video features along the time axis. The ReLU activation function sets negative similarities to zero, enhancing robustness and reducing the impact of noise.Since \( \bar{d}_i \in [0, 1] \) allowing for the direct computation of the binary cross-entropy loss (BCE).

\subsubsection{\textbf{Generative synchronization  }}

In generative synchronization,we utilize a frozen pre-trained ASR model as the Speech Recognition (SR) head of CoGenAV to generate speech text. The core objective is to enforce the feature representations output by CoGenAV to simultaneously align with both the \textbf{acoustic representations }and \textbf{text-semantic representations} of the pre-trained ASR model in the latent space. This ensures that the CoGenAV representations encode both acoustically correlated information from lip movements (e.g., phoneme articulation) and text-semantic information consistent with the speech signal, while directly leveraging the ASR model's acoustic-to-text decoding capabilities.

To mitigate the frame rate discrepancy and modality misalignment between CoGenAV and the SR head, we  a lightweight  \textbf{Feature Adaptation Module}, which comprises a Delta Upsampler and a GatedFFN-MHA Layer.

The Delta Upsampler is designed to bridge  frame-rate gap between CoGenAV's features (25 fps) and the frozen SR head (which expects 50 fps acoustic features). This component uses temporal differential convolution layers and leverages dual-channel convolution outputs to construct odd/even interpolated frames (a novel design inspired by \cite{TDN}) to explicitly model frame-to-frame changes and interpolate features, effectively doubling the temporal resolution from \( T \) to \( 2T \). This approach mitigates the semantic abruptness and high-frequency information loss observed in adjacent interpolated frames caused by hard-coded replication schemes \cite{whisperer}.

The GatedFFN-MHA Layer further addresses the modality misalignment between  CoGenAV's features and the pre-trained ASR model, reducing training difficulty and accelerating convergence. It employs multi-head self-attention (MHA) to capture the temporal context within the adapted visual stream, followed by a gated feed-forward network (GatedFFN). The GatedFFN (Eq.\ref{eq:gate_ffn}) incorporates a gating mechanism similar to that of Gated Linear Units \cite{GLU} to selectively modulate the feature representation, potentially filtering out modality-irrelevant noise before passing the features to the frozen SR head, thereby facilitating cross-modal adaptation.
\begin{gather}  
x' = x + \sigma(\text{FFN}(x)) \cdot \text{FFN}(x) \label{eq:gate_ffn}  
\end{gather}  
For the generative loss \(\mathcal{L}_{\text{Gen}}\), we adopt the same supervision strategy as the pre-trained ASR model, which maximizes the log-likelihood of ground-truth text transcriptions(Eq.\ref{eq:L_gen}):  
\begin{gather}  
  \mathcal{L}_{\text{Gen}} = -\mathbb{E}_{(x, s^*) \sim \mathcal{D}} \log p(s^* \mid x)  
  \label{eq:L_gen}  
\end{gather}
where \(x\) denotes input features, \(s^*\) is the target text sequence, and \(\mathcal{D}\) represents the training dataset.

\subsubsection{\textbf{Training detail}}

During training, we employ Contrastive-Generative synchronization with joint supervision, leading to an overall loss defined as(Eq.\ref{eq:L_sum}):  
\begin{gather}  
  \mathcal{L} = \mathcal{L}_{\text{Gen}} + \lambda \mathcal{L}_{\text{Co}}
  \label{eq:L_sum}  
\end{gather}
where \(\lambda\) is a balancing hyperparameter.  

To encourage CoGenAV to learn more representative features and reduce dependence on audio information, we implement a modality dropping strategy for the features input to the pre-trained ASR model. Based on our experiments(Talel.\ref{tab:ablation}), we selected the proportions of pure audio, pure visual, and audio-visual data to be 20\%, 40\%, and 40\%, respectively. This stochastic approach enhances robustness and ensures balanced multimodal representation learning.

\subsection{\textbf{CoGeneAV on Audio-Visual Speech-Centric task }}
CogenAV employs a contrastive-generative synchronization framework to effectively leverage cross-modal correlations among audio, visual, and textual data. The generated modality-aligned speech representations exhibit high adaptability and can be effectively applied to various speech-centric scenarios involving human speech videos, including but not limited to: Visual Speech Recognition (VSR), Audio-Visual Speech Recognition (AVSR), Audio-Visual  speech Enhancement(AVSE), Audio-Visual speech separation(AVSS), and Active Speaker Detection (ASD).

\subsubsection{\textbf{Visual and Audio-Visual  speech recognition (VSR/AVSR)}}

Visual Speech Recognition (VSR) involves recognizing speech text solely by analyzing lip movement features. CoGenAV directly serves as a complete VSR model by inputting only visual features without architectural modifications. The pretrained visual encoder captures fine-grained lip dynamics, enabling accurate text prediction solely from lip movements.  

Audio-Visual Speech Recognition (AVSR) enhances speech recognition by jointly modeling audio signals and lip movements, CoGenAV can similarly be applied by inputting audio and visual streams in a baseline setup, where fused multimodal features leverage synchronized cross-modal cues for improved recognition. We also propose a novel adaptation strategy for clean speech AVSR,As shown in  Figure~\ref{fig:archmodel} (a)(right) . CoGenAV’s upsampled visual features (Key and Value) interact with Whisper-extracted audio features (Query) via a cross-attention mechanism, replacing the original self-attention in the GatedFFN-MHA layer. The Whisper model remains frozen, while only the lightweight Feature Adaptation module (10M parameters) is trained for efficient audio-visual fusion.  

\subsubsection{\textbf{Audio-Visual Speech separation and Enhancement(AVSS/AVSE)}}

Audio-Visual Speech Separation(AVSS) aims to isolate individual speaker voices from mixed audio using visual cues (lip movements), while Audio-Visual Speech Enhancement(AVSE) focuses on denoising speech by suppressing background interference. We treat AVSE as a specialized AVSS case where noise is separated from the target speaker.  

As shown in Figure~\ref{fig:archmodel}(b)(right), frozen CoGenAV visual features (768D) are integrated into AV-SepFormer \cite{av-sepformer}, replacing its default 512D visual encoder. This adapts the separation head to leverage CoGenAV’s articulatory-aware representations for disentangling target speech from noise or competing speakers. The modified architecture retains AV-SepFormer’s core separation mechanism while aligning visual feature dimensions to CoGenAV’s output space.

\subsubsection{\textbf{Active speaker detection (ASD)}}

Active Speaker Detection (ASD) aims to identify the active speaker in multi-person scenarios by verifying the synchronization between lip movements and corresponding speech signals.  
CoGenAV enhances ASD by replacing the original audio and visual encoders in LRASD \cite{LRASD} with its synchronized multimodal representations,As shown in Figure~\ref{fig:archmodel}(c)(right). These pretrained features inherently encode temporal alignment cues (audio-visual correspondence) and semantic consistency, enabling robust speaker verification through cross-modal feature matching. 

\section{Experimental setup}

\subsection{Datasets }

In most of the tasks described in this paper, we primarily utilize the LRS2~\cite{LRS2} dataset over others (e.g., LRS3~\cite{LRS3}) as it is currently the largest publicly available English audio-visual speech recognition resource under standard academic licensing. LRS2 was collected from BBC programs and contains a total of 144,482 video clips, amounting to 223 hours of content. During the training of CoGenAV, we randomly added "natural," "music," and "babble" noise from the MUSAN dataset \cite{MUSAN} to the speech audio of LRS2 at a signal-to-noise ratio (SNR) ranging from -5 dB to 5 dB to enhance data diversity. In particular, for the Speech Separation task, we mixed audio from two different speakers in the dataset within the SNR range of -5 dB to 5 dB to create a dataset with overlapping speech. For the multi-speaker Active Speaker Detection (ASD) task, we used the Talkies \cite{Talkiez} dataset, which contains 10,000 video clips and is annotated with 23,507 facial trajectories, averaging 2.35 facial trajectories per video.

\subsection{Data processing}

For the visual stream, we followed the preprocessing steps outlined in previous work \cite{Auto-avsr}. Based on the preprocessing code from \cite{Auto-avsr}, we extracted facial keypoint information to identify the mouth region, after which we cropped the region of interest (ROI) using a bounding box of size \(96 \times 96\). The frame rate of all videos was standardized to 25 fps. 
For the audio stream, we adopted the same preprocessing scheme as Whisper \cite{whisper}, extracting 80 bin log Mel spectrograms from audio sampled at 16 kHz, with a stride of 10 ms and a window size of 25 ms.

\subsection{Implementation details}

In terms of model architecture, we consider two model configurations: Base and Large.The main difference is that the Base configuration has 12 transformer blocks with a feature dimension of 768, while the Large configuration has 24 transformer blocks with a feature dimension of 1024. The trainable parameters for the Base model are 120M, while the trainable parameters for the Large model are 360M.By the way, when training the Base model, we used the frozen pre-trained ASR model Whisper-small, while for training the Large model, we chose Whisper-medium.

Regarding data augmentation, we applied horizontal flipping and random cropping to the visual inputs, while adding random noise to the audio stream.For weight initialization, we initialized the visual encoder and transformer encoder with AV-HuBERT weights, while the audio encoder and Feature Adaptation module were initialized randomly. Regarding the training strategy, we initially fine-tuned the randomly initialized components before performing global training of CoGenAV. Although the first phase did not achieve strong performance, it significantly accelerated the model's overall convergence speed.Regarding model testing, in the VSR and AVSR tasks, we set the beam size to 3.In terms of training parameters, we used the Adam optimizer with 500 warmup steps and an initial learning rate of 1e-5.We trained the model on two A100 GPUs, with a batch size of 16 per GPU and 4 gradient accumulation steps.

\section{Resluts}

\subsection{CoGenAV for VSR }

\begin{table}
\centering
\vspace{1em} 
\caption{\label{tab:vsr} The WER (\%) of our VSR models compared to previous works trained exclusively on the LRS2 dataset. V=visual.}
\vspace{1em} 
\begin{tabular}{c c c c}
\toprule 
Method & Modalities & Label Hours & WER (VSR \%) \\ 
\midrule 
CTC/Attention~\cite{ctc_attentin} & V & 380 & 63.5 \\ 
AV-HuBERT Base~\cite{av-hubert}~\cite{VatLM} & V & 223 & 31.2 \\ 
ES³ Base~\cite{ES3} & V & 223 & 29.3 \\ 
SyncVSR~\cite{SyncVSR} & V & 223 & 28.9 \\ 
VTP~\cite{VTP} & V & 698 & 28.9 \\ 
AutoAVSR~\cite{Auto-avsr} & V & 818 & 27.9 \\ 
Whisperer Large~\cite{whisperer}& V & 223 & 26.3 \\ 
\midrule 
CoGenAV Base (ours) & V & 223 & \textbf{24.8} \\ 
CoGenAV Large (ours) & V & 223 & \textbf{20.5}\\ 
\bottomrule 
\end{tabular}
\vspace{1em} 
\end{table}

In Table~\ref{tab:vsr}, we present a comparative analysis of the performance of the CoGenAV approach on the LRS2 test set alongside previous works.  Considering that an increase in training data directly enhances the performance metrics of VSR models \cite{Auto-avsr,USR,SyncVSR}, we primarily compared the metrics based on training data on LRS2. Experimental results indicate that our CoGenAV Base model significantly outperformed other models, achieving a leading word error rate (WER) of 24.8, which is an improvement of 6.4 over the AV-HuBERT Base reported in \cite{VatLM}. The CoGenAV Large model achieved even better results, with a word error rate of 20.5, along with a 7.4\% improvement over AutoAVSR \cite{Auto-avsr} which was trained on 818 hours, and an 8.4\% improvement over SyncVSR \cite{SyncVSR} which was trained on 223 hours.These results indicate that our model demonstrates a significant advantage with a small amount of training data.Moreover, our CoGenAV Large outperforming the Whisperer \cite{whisperer} Large(which employed the VSR methods of AV-HuBERT Large and Whisper Medium, results we reproduced in our own experiments) by 5.8, indicating that the features generated by our CoGenAV perform better than those of AV-HuBERT.

\subsection{CoGenAV for AVSR }
\subsubsection{Noise results }

Recognizing speech text in noisy environments remains a challenging task. In the LRS2 test set, we introduced noise with a signal-to-noise ratio (SNR) of 0 dB, revealing a significant degradation in the recognition capabilities of the Whisper model. For example, the word error rate (WER) for Whisper-Medium increased from 6.4 to 34.2 after adding noise. To explore whether this performance decline stemmed from a lack of noise data during Whisper's training, we fine-tuned the Whisper-Medium model using data that included noise signals. However, this only marginally improved performance, achieving a WER of 20.6. This suggests that although incorporating noise during training is beneficial, the overlapping of noise with the speech signal spectrum poses considerable challenges for the model in recognizing speech text.

In contrast, during the training of CoGenAV, we jointly modeled noisy acoustic signals and lip movement features using Seq2Seq Contrastive loss, effectively aligning audio and visual features across modalities. This integration allowed CoGenAV to leverage rich temporal and semantic correlations, resulting in robustness to noise in the AVSR task.Our results(Table~\ref{tab:avsr_noise}) demonstrate that CoGenAV Large achieved a score of 2.6 in noisy environments, improving 18.0 points compared to the fine-tuned score of 20.6 for Whisper-Medium on the same noisy data—an enhancement of over 70\%. Notably, Whisper-Medium served as the SR Head, with its model weights frozen, while the original model had a WER of 34.2 in noisy conditions. This indicates that CoGenAV extracted higher-quality features compared to the original Whisper Audio Layer, resulting in improved performance metrics. The same robust performance was observed for CoGenAV Base in noisy environments.

\begin{table}
\centering
\vspace{1em} 
\caption{\label{tab:avsr_noise}The WER (\%) of our models on LRS2 with babble noise injected at a 0-SNR level (noisy). A=audio, AV=audio-visual. "0" indicates that the Whisper model remains frozen, "*" denotes the Whisper model fine-tuned on LRS2.}
\vspace{1em} 
\begin{tabular}{c c c c}
\toprule 
Method & Modalities & Base& Large\\ 
\midrule 
Whisper\(^0\) & A & 43.8& 34.2\\ 
Whisper\(^*\) & A & 24.8 & 20.6 \\ 
\midrule 
CoGenAV+Whisper\(^0\) (ours) & AV & \textbf{5.2}& \textbf{2.6}\\ 
\bottomrule 
\end{tabular}
\vspace{1em} 
\end{table}

\subsubsection{Clean results }

For AVSR in the case of clean voice, we adapt CoGenAV’s pretrained visual features (Key and Value) to interact with frozen Whisper audio features (Query) via cross-attention (Figure~\ref{fig:archmodel}(a)(right)). Only the Feature Adaptation module(10M parameters) is fine-tuned, while the Whisper model and CoGenAV backbone remain frozen, ensuring efficient training with minimal computational overhead.

As shown in Table~\ref{tab:avsrlrs2},using the original weights of Whisper-Medium as the SR Head for CoGenAV during text generation, our WER improved from 6.4 to 1.8. When using the Whisper-Medium fine-tuned on clean audio from LRS2 as CoGenAV's SR Head, the WER decreased from 1.5 to 1.27, establishing a new state-of-the-art (SOTA) result on the LRS2 dataset. 
As Table~\ref{tab:avsrlrs2} illustrates, CoGenAV establishes a new state-of-the-art AVSR result on LRS2 (\textbf{1.27\% WER}), outperforming prior works like Auto AVSR \cite{Auto-avsr} (\textbf{1.5\% WER}) while using \textbf{15\(\times\) less training data} (223 vs. 3,448 hours). Notably, our method surpasses all existing models in the "clean audio" setting, including recent multimodal architectures (e.g., USR \cite{USR}: 1.9\% WER) and Whisper variants, despite their reliance on larger datasets or full-model fine-tuning.These results validate that CoGenAV's contrastive-generative framework not only enables efficient cross-modal fusion but also extracts exceptionally discriminative visual representations, achieving unprecedented accuracy with minimal labeled data.
\begin{table}
\centering
\vspace{1em} 
\caption{\label{tab:avsrlrs2} The WER (\%) of our  CoGenAV Large and comparisons with previous works presented on the LRS2 test set. A=audio, AV=audio-visual. "0" indicates that the Whisper model remains frozen, "*" denotes the Whisper model fine-tuned on LRS2.}
\vspace{1em} 
\begin{tabular}{c c c c}
\toprule 
Method & Modalities & Label Hours & WER (\%) \\ 
\midrule 
CTC/Attention~\cite{ctc_attentin} & AV & 380 & 7.0 \\ 
CM-seq2seq~\cite{CM2SEQ} & AV & 380 & 3.7 \\ 
MoCo+Wav2Vec~\cite{MoCo+Wav2Vec} & AV & 223 & 2.6 \\ 
Efficient Conformer~\cite{EfficientAVSR} & AV & 223 & 2.3 \\ 
USR~\cite{USR} & AV & 223 & 1.9 \\ 
AutoAVSR~\cite{Auto-avsr} & AV & 818 & 2.6 \\ 
AutoAVSR~\cite{Auto-avsr} & AV & 3448 & 1.5 \\
 Whisper-flamingo~\cite{Whisper-flamingo}& AV & 223&1.4\\\midrule
 Whisper\(^0\)& A& -&6.4\\
 CoGenAV+Whisper\(^0\)(ours) & AV & 223 &\textbf{1.8} \\ 
\midrule 
Whisper\(^*\)& A & 223 & 1.5 \\ 
CoGenAV+Whisper\(^*\)(ours) & AV & 223 & \textbf{1.27} \\ 
\bottomrule 
\end{tabular}
\vspace{1em} 
\end{table}

\subsection{CoGenAV for AVSS}

While prior work (e.g. AV-HuBERT \cite{avhubert-se}) demonstrates the applicability of audio-visual features to tasks like speech enhancement (AVSE) and separation (AVSS), CoGenAV possesses more effective and versatile  representations, achieves superior generalization across tasks without task-specific architectural modifications.

The method of applying CoGenAV to the Speech Separation task is illustrated in Figure \ref{fig:archmodel}(b)(right). We extract the speaker's visual features using the frozen CoGenAV Base model, and then use AV-SepFormer \cite{av-sepformer} as the Speech Separation Head. 

We adopted the same data setting as AvSepChain~\cite{AvSepChain}. In Table~\ref{tab:avss}, we compare CoGenAV with some previous methods, where CoGenAV achieves a leading result with an SDRi of 16.0 dB. When using the same Speech Separation Head, CoGenAV as the Visual Encoder improves upon Av-HuBERT in the AVSS task by 1.6 dB. Similarly, our SDR metric is also better than that of AvSepChain~\cite{AvSepChain} by 0.3 dB, which is a model that applies Av-HuBERT to solve the AVSS task.

\begin{table}
\centering
\vspace{1em} 
\caption{\label{tab:avss} Audio-Visual Speech separation results on LRS2. These metrics represent the average values for all speakers in each test set, where larger SI-SNRi, SDRi, PESQ are better.}
\vspace{1em} 
\begin{tabular}{c c c c c}
\toprule 
Method & Modalities & SI-SNRi & SDRi & PESQ \\ 
\midrule 
AVConvTasNet~\cite{AVConvTasNet} & AV & 12.4 & 12.7 & 2.75 \\ 
VisualVoice~\cite{VisualVoice} & AV & 11.5 & 11.8 & 3.0 \\ 
MUSE~\cite{Muse} & AV & 13.5 & 13.8 & 2.97 \\ 
CTCNet~\cite{CTCNet} & AV & 14.3 & 14.6 & 3.06 \\ 
AvSepChain~\cite{AvSepChain} & AV& 15.3 & 15.7 & \textbf{3.26} \\
\midrule 
 Av-SepFormer~\cite{av-sepformer}& AV& 14.1& 14.4&3.15\\ 
CoGenAV (ours) & AV & \textbf{15.7} & \textbf{16.0} & 3.23 \\ 
\bottomrule 
\end{tabular}
\vspace{1em} 
\end{table}

\subsection{CoGenAV for AVSE}
\begin{table}
\centering
\vspace{1em} 
\caption{\label{tab:avse}Audio-Visual Speech Enhancement results on LRS2. These metrics represent the average values for all speakers in each test set, where larger SI-SNRi, SDRi, PESQ are better.}
\vspace{1em} 
\begin{tabular}{c c c c c}
\toprule 
Method & Modalities & SI-SNRi & SDRi & PESQ \\ 
\midrule 
AV-HuBERT-SE~\cite{avhubert-se}& AV & - & - & 1.40 \\ \midrule 
Av-SepFormer~\cite{av-sepformer} & AV & 6.6 & 7.4 & 2.46 \\ 
CoGenAV (ours) & AV & \textbf{8.3} & \textbf{9.0} & \textbf{2.56}\\ 
\bottomrule 
\end{tabular}
\vspace{1em} 
\end{table}

The speech enhancement task AVSE and audio-visual speech separation AVSS are similar. For AVSE, we treat it as a special case of the AVSS task, specifically separating noise from the speaker. We employ the same method as in AVSS, as shown inFigure \ref{fig:archmodel}(b)(right). During training, we dynamically add noise to the audio in LRS2, and we add fixed strong noise at 0 dB in the test data. In Table~\ref{tab:avse}, we compare the performance of CoGenAV with Av-HuBERT in the AVSE task, where CoGenAV achieves a leading result with an SDRi of 9.0 dB, surpassing Av-HuBERT by 1.6 dB.

\subsection{CoGenAV for ASD }
\begin{table}
\centering
\vspace{1em} 
\caption{\label{tab:muti-speaker} The mAP metric for ASD on the Talkies.}
\vspace{1em} 
\begin{tabular}{c c c}
\toprule 
Method & Modalities & mAP@ASD \\ 
\midrule 
Mass~\cite{Talkiez} & AV & 79.7 \\ 
EASEE~\cite{EASEE} & AV & 93.6 \\ 
Light-ASD~\cite{light_asd} & AV & 93.9 \\ 
LocoNet~\cite{LoCoNet} & AV & 96.1 \\ 
\midrule 
CoGenAV (ours) & AV & \textbf{96.3} \\ 
\bottomrule 
\end{tabular}
\vspace{1em} 
\end{table}

CoGenAV’s synchronized audio-visual representations are inherently suited for ASD, which requires precise temporal alignment between lip movements and speech. As shown in Figure \ref{fig:archmodel}(c)(right), we replace the audio and visual encoders in LRASD \cite{LRASD} with frozen CoGenAV features, leveraging its pretrained cross-modal synchronization cues to detect active speakers. This approach eliminates the need for task-specific training of modality encoders.The input videos for CoGenAV require preprocessing adapted from the AutoAVSR ~\cite{Auto-avsr} method.

As shown in Table~\ref{tab:muti-speaker}, we compare the performance of CoGenAV with previous methods on the Talkies dataset \cite{Talkiez}, where our model achieves an average mean accuracy (mAP) of 96.3\%, surpassing prior work \cite{Talkiez,EASEE,light_asd,LoCoNet}. 

\section{Ablation Study  and Analysis }

\subsection{Ablation Study}
To evaluate the contribution of key components of CoGenAV, we conducted ablation studies on the LRS2 VSR and AVSR task using the CoGenAV Large configuration. Results are shown in Table \ref{tab:ablation} and  Table \ref{tab:ablation_adaptaion}.

\textbf{Impact of Training Objectives}
On VSR task,Training CoGenAV with only the generative synchronization loss \(\mathcal{L}_{\text{Gen}}\) resulted in a WER of 22.5. When varying the balancing hyperparameter \( \lambda \) in the combined loss, we found optimal performance around \( \lambda = 1 \) (WER: 20.4), confirming the significant benefit of our dual contrastive-generative objective over using either loss in isolation. In terms of the training proportions of different modal data, we also conducted simple tests and found that increasing the proportion of the Audio-Visual modality may slightly reduce the VSR metrics, but significantly improves the AVSR metrics.

\begin{figure}
\centering
\includegraphics[width=0.75 \linewidth]{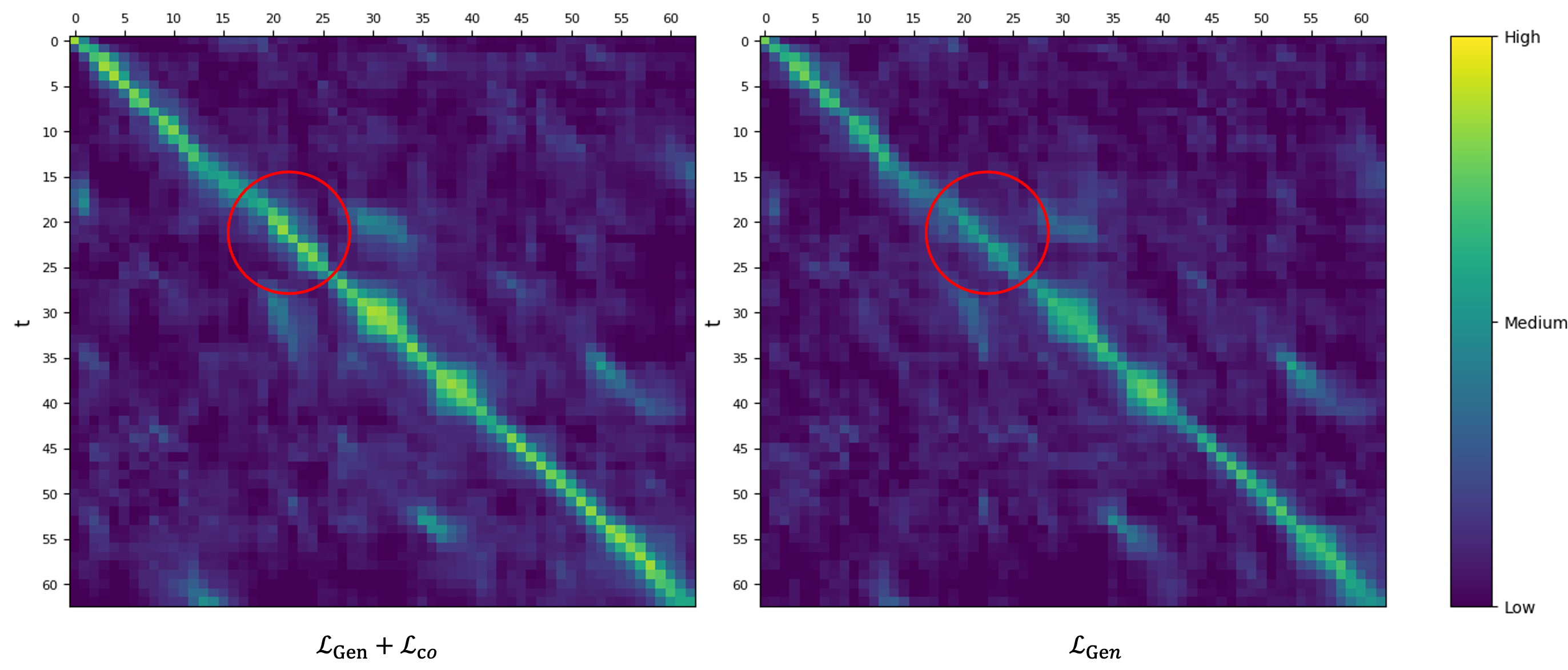}
\caption{\label{fig:heatmap}
Cross-Modal Alignment Heatmap. The brighter the color, the higher the similarity between audio-visual features, indicating better alignment. Left: \(\mathcal{L}_{\text{Gen}}\) with \(\mathcal{L}_{\text{Co}}\); Right: only \(\mathcal{L}_{\text{Gen}}\) without \(\mathcal{L}_{\text{Co}}\).}
\end{figure}

To intuitively analyze the synchronization between audio and visual modalities at different time points with and without \(\mathcal{L}_{\text{Co}}\), we construct CoGenAV cross-modal alignment heatmaps for both scenarios (Fig. \ref{fig:heatmap}). Using \textbf{CoGenAV}, we extract audio and visual features from a synchronized lip-reading video and compute frame-wise cosine similarity between audio and video frames to generate a similarity matrix. Each element in this matrix represents the similarity level between audio and visual features at corresponding time points. The heatmap is then plotted based on this matrix, where brighter regions indicate higher similarity and darker regions indicate lower similarity. Ideally, the main diagonal of the heatmap should be prominently highlighted, indicating strong temporal alignment between audio and visual features.
\begin{table}
\centering
\vspace{1em} 
\caption{\label{tab:ablation} Ablation analysis for \( \lambda \mathcal{L}_{\text{Co} }\) and modality dropping.  }
\vspace{1em} 
\begin{tabular}{c c cl}
\toprule 
 \(\lambda\) of \(\mathcal{L}_{\text{Co}}\)&  modality dropping& VSR &AVSR(Nosie) \\ \midrule 
 0& AV:0.2 V:0.6 A:0.2& 22.5&  3.26\\
 1& AV:0.2 V:0.6 A:0.2& 20.4&3.0 \\
 1& AV:0.4 V:0.4 A:0.2& 20.5& 2.6\\
 2& AV:0.4 V:0.4 A:0.2& 21.2&2.6\\ 
 \midrule 
\end{tabular}
\vspace{1em} 
\end{table}
Overall, regardless of the use of \(\mathcal{L}_{\text{Co}}\), the diagonal of our heatmap is notably bright, which demonstrates that our generative synchronization achieves good alignment even when only using \(\mathcal{L}_{\text{Gen}}\). In the case use \(\mathcal{L}_{\text{Gen}}\) with \(\mathcal{L}_{\text{Co}}\) (Fig. \ref{fig:heatmap} left), the diagonal region is even brighter than when only using \(\mathcal{L}_{\text{Gen}}\) (Fig. \ref{fig:heatmap} right), indicating that the joint training of \(\mathcal{L}_{\text{Co}}\) and \(\mathcal{L}_{\text{Gen}}\) further enhances the alignment between audio and visual features. This results in stronger adaptability for downstream tasks that require audio-visual aligned features (e.g., audio-visual synchronization tasks, ASD, etc.). This clearly illustrates the effectiveness of our CoGenAV in cross-modal alignment tasks.

\textbf{Impact of Feature Adaptation Module}
In the same training strategy experimental setting, we compared the impact of various components of the Feature Adaptation Module on the VSR and AVSR with Noise tasks. Results are shown in Table \ref{tab:ablation_adaptaion}.In the VSR task, removing the entire feature adaptation module and directly feeding CoGenAV features into the SR head (only using simple repetition upsampling) severely degrades performance, resulting in a VSR of 27.5. Replacing our GateFFN with a standard FFN leads to a drop in the metric from 20.4 to 22.2, a decrease of 1.7. Substituting our Delta Upsampler with simple repetition upsampling results in a 0.6 decrease in the metric. Similar conclusions can be drawn for the AVSR with Noise scenario. These results validate the effectiveness of our proposed Feature Adaptation Module.
\begin{table}
\centering
\vspace{1em} 

\caption{\label{tab:ablation_adaptaion} Ablation analysis for  Feature Adaptation Module.}
\vspace{1em} 
\begin{tabular}{c c cl}
 & & &\\
\toprule 
  Feature Adaptation&  Deta Upsampler& VSR &AVSR(Nosie) \\
  \midrule 
  -& \(\times\) & 27.5&7.1\\ 
  FFN+MHA& \(\checkmark\)& 22.2&3.14\\
  Gate FFN+MHA& \(\times\)& 21.1&3.0\\ 
  Gate FFN+MHA& \(\checkmark\)& 20.4& 3.0\\ \midrule 
\end{tabular}
\vspace{1em} 
\end{table}

\subsection{Limitations and Future Work}
Although CoGenAV achieves state-of-the-art results on multiple audio-visual tasks, its evaluations are primarily conducted on the LRS2\cite{LRS2} dataset. More comprehensive testing could be performed once LRS3\cite{LRS3}  becomes publicly accessible in the future.

\section{Conclusion}
This paper presents CoGenAV, a unified contrastive-generative framework that establishes tri-modal alignment representations across audio (A), visual (V), and text (T) streams for universal adaptation in speech-centric applications. We present a novel audio-visual sequence-to-sequence contrastive learning approach that effectively captures precise temporal alignment between modalities. Extensive experiments validate the versatility and data efficiency of CoGenAV, achieving state-of-the-art performance across various audio-visual tasks with only moderate training data.

\clearpage  
{
\medskip
\small  
  
}

\end{document}